\documentclass [aps,prb,amssymb,amsmath,twocolumn,showpacs,superscriptaddress]{revtex4-2}

\usepackage{microtype}
\usepackage{amsmath}
\usepackage{amssymb}
\usepackage{graphicx}
\usepackage{verbatim}
\usepackage{bm}
\usepackage{physics}
\usepackage{mathrsfs}
\usepackage{pifont}
\usepackage{gensymb}
\usepackage{graphicx,epsfig,amsfonts}
\usepackage{hyperref}
\usepackage{xcolor}
\hypersetup{colorlinks=true,linkcolor=blue,citecolor=blue,urlcolor=blue}
\usepackage{times}
\usepackage{lipsum}
\usepackage[all]{xy}
\usepackage{soul}
\soulregister{\cite}7
\soulregister{\eqref}7
\soulregister{\ref}7
\soulregister{\bibitem}7

\begin{document}

\title{Two-dimensional Weyl nodal-line semimetal and antihelical edge states in a modified Kane-Mele model}

\author{Xiaokang Dai}
\affiliation{School of Physics and Electronics, Hunan University, Changsha 410082, China}

\author{Pei-Hao Fu}
\affiliation{Science, Mathematics and Technology, Singapore University of Technology and Design, Singapore 487372, Singapore}

\author{Yee Sin Ang}
\affiliation{Science, Mathematics and Technology, Singapore University of Technology and Design, Singapore 487372, Singapore}

\author{Qinjun Chen}
\email{chenqj@hnu.edu.cn}
\affiliation{School of Physics and Electronics, Hunan University, Changsha 410082, China}

\date{\today}

\begin{abstract}
The Kane-Mele model has been modified to achieve versatile topological phases. Previous work [\href{https://journals.aps.org/prl/abstract/10.1103/PhysRevLett.120.156402}{Phys. Rev. Lett. \textbf{120}, 156402 (2018)}] introduced a staggered intrinsic spin-orbit coupling effect to generate pseudohelical edge states, with Rashba spin-orbit coupling facilitating spin flips in alternating sublattices. 
Our study demonstrates that, in the absence of Rashba spin-orbit coupling, the modified Kane-Mele model with staggered intrinsic spin-orbit coupling evolves into a $Z_{2}$ class topological metal, specifically a two-dimensional Weyl nodal-line semimetal. 
In a nanoribbon geometry, we predict the emergence of antihelical edge states, which support spin-polarized currents flowing in the same direction along parallel boundaries. 
Unlike pseudohelical edge states, antihelical edge states can be viewed as a superposition of two antichiral edge states related by time-reversal symmetry. 
However, the spin Hall conductance from antihelical edge states is not quantized due to the presence of gapless bulk states. 
Additionally, we examine the robustness of helical, pseudohelical, and antihelical edge states in the presence of nonmagnetic disorders, highlighting the particular fragility of antihelical edge states. 
Our findings enhance the understanding of the modified Kane-Mele model, providing new insights into its topological properties.
\end{abstract}
\maketitle

\section{\label{sec:Introduction}Introduction}
A large number of quantum materials have been discovered that exhibit extraordinary edge states originating from their bulk topological properties \cite{Zhang2019,Feng2019,Vergniory2019,Vergniory2022}. 
Two key indicators of these topological phases are the gap closing and opening during the topological phase transitions and the entanglement of the wave functions \cite{Keimer2017} that define the nontrivial quantum invariants. 
Typically, the critical gap closing conditions, derived from the Hamiltonian under periodic boundary conditions, delineate the phase boundaries that construct the phase diagram \cite{Ezawa2013,Chiu2016,Dai2024}, whereas the topological invariants provide insight into the mechanisms of the bulk-boundary correspondence \cite{Mizoguchi2021}. 
Notable examples include the Haldane model \cite{Haldane1988} and the Kane-Mele (KM) model \cite{Kane20051}, which have been significantly modified to predict intriguing nontrivial topological edge states \cite{Shevtsov2012,Colomes2018,Frank2018,Wang2021,Zhuang2022,Yu2022}.

In the Haldane model with broken time-reversal (TR) symmetry, edge modes achieve the chirality due to periodic local magnetic-flux density and different next-nearest-neighbor (NNN) hopping terms in different sublattices \cite{Haldane1988}. In a strip geometry, these chiral edge modes propagate in opposite directions along parallel boundaries \cite{Hasan2010,Qi2011}. 
A subtle modification to the Haldane term, as proposed by Colom{\'e}s \textit{et al.} \cite{Colomes2018}, allows the NNN hopping term to act equally in both \textit{A} and \textit{B} sublattices, resulting in a scalar potential with an opposite sign in each valley. 
Consequently, each boundary of a strip geometry supports an edge mode propagating in the same direction at opposite boundaries. 
These edge states are known as the antichiral edge states \cite{Mandal2019,Zhou2020,Bhowmick2020,Chen2020}, which are topologically protected with gapless bulk supporting compensation modes with opposite velocities. Pictorially, this can be understood as the “topological coaxial cable” scheme \cite{Pikulin2016,Schuster2016}, which demonstrates the spatial separation of bulk and edge modes, thus suppressing the backscattering.

\begin{figure}
    \centering
    \includegraphics[width=3 in]{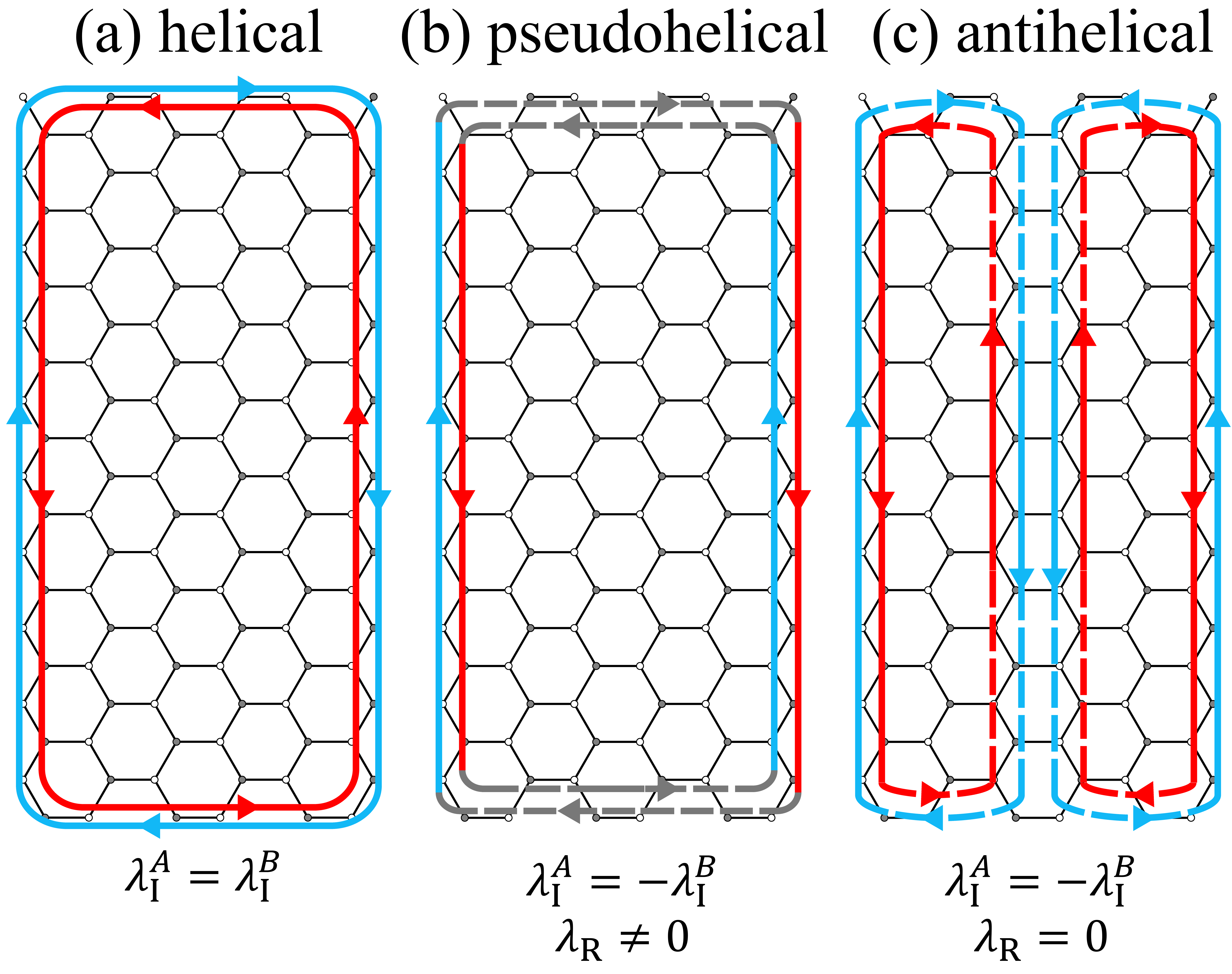}
    \caption{Schematic representations of (a) helical, (b) pseudohelical, and (c) antihelical edge states. The lines with arrows colored in red (light blue) denote spin-up (spin-down) edge states. The gray dashed lines in (b) represent the perfect tunneling along the armchair boundaries due to Rashba SOC. The red and light blue dashed lines in (c) denote the spin-polarized bulk states.}
    \label{Fig1}
\end{figure}

In systems where TR symmetry is restored, such as the KM model \cite{Kane20051} or the Bernevig-Hughes-Zhang model \cite{Bernevig2006}, spin conservation and NNN spin-orbit coupling (SOC) lead to the prohibition of an odd number of chiral edge states at each boundary, as claimed by the Kramers theorem \cite{Wu2006}. 
In this scenario, the topological edge states become helical, featuring spin-momentum locking, as sketched in Fig. \ref{Fig1}(a). 
The helical edge states are generally regarded as two copies of chiral edge state \cite{Kane20051,Kane20052}, with the same spin-polarized edge modes propagating in opposite directions along distinct boundaries. 
An analogous modification in the KM model by replacing the NNN intrinsic SOC with a staggered NNN intrinsic SOC \cite{Frank2018,Wang2015,Gmitra2015,Gmitra2016}, leads to a quantum valley spin Hall state with vanishing $Z_{2}$ but symmetry-protected pseudohelical edge states (PHESs) as illustrated in Fig. \ref{Fig1}(b) \cite{Frank2018}. 
These PHESs feature spin-polarized currents propagating co-directionally at opposite boundaries, with Rashba SOC causing the bulk to gap and allowing perfect tunneling along the armchair boundaries due to the spin flips.

In the absence of Rashba SOC, the modified KM (MKM) model can be decoupled into two copies of the modified Haldane model with opposite flux \cite{Colomes2018}. 
This results in a closed bulk gap, describing a two-dimensional (2D) Weyl nodal-line semimetal (NLS) with broken inversion symmetry \cite{Young2015,Lv2021}. 
While the edge states in the Weyl NLS phase [see Fig. \ref{Fig1}(c)] resemble PHESs, they originate from a different bulk-boundary correspondence. 
As depicted in Fig. \ref{Fig1}(c), in the NLS phase, bulk states compensate for the edge spin currents, suggesting that these edge states should be distinguished from PHESs with Rashba SOC acting as an order parameter for the topological phase transition between these states. 
Therefore, the edge states in the Weyl NLS phase are termed as antihelical edge states (AHESs), analogous to antichiral edge states.
Notably, the edge states expected in 2D NLS connect the nodal lines in two different valleys, which is distinct from the three-dimensional NLSs, where the drumhead surface states are enclosed by the nodal lines \cite{Fu2020}.
Additionally, akin to another type of 2D Weyl NLS phase with broken TR symmetry \cite{Liu2018,Lv2022,Li2024}, the nontrivial topology implies that the NLS phase can be regarded as the parent phase of the quantum anomalous Hall insulator \cite{Hogl2020,Vila2021} and quantum valley Hall insulator phases \cite{Ezawa2013,Wang2015,Alsharari2016,Khatibi2023}.

In this paper, we carefully investigate AHESs in an MKM in the absence of Rashba SOC, as well as the Weyl NLS phases that are associated with these edge states. The rest of the paper is arranged as follow.
In Sec. \ref{sec:Model and Phase Diagram}, we establish a phase diagram for the MKM model by examining the band gap at threefold-symmetry invariant points, uncovering Weyl NLS and quantum spin Hall insulator (QSHI) phases distinguished by bulk band features and spin Hall conductance. 
Section \ref{sec:Edge States} presents the band structure of a zigzag-terminated MKM nanoribbon and the analytical expressions for the energy dispersion of edge states, delving into their characteristics and evolutions during the phase transitions. 
We find that in the NLS phase, the AHESs exhibit perfect reflection at the armchair boundaries. 
We calculate the transport properties of topological nanoribbons in the presence of nonmagnetic impurities and discuss the disorder-induced scattering mechanism of three types of edge states in Sec. \ref{sec:Robustness of Edge States}. 
Finally, Sec. \ref{sec:Conclusion and Discussion} summarizes the main results of this paper and discusses the potential experimental realization of AHESs.

\section{\label{sec:Model and Phase Diagram}Model and Phase Diagram}
\begin{figure*}
    \centering
    \includegraphics[width=5.2 in]{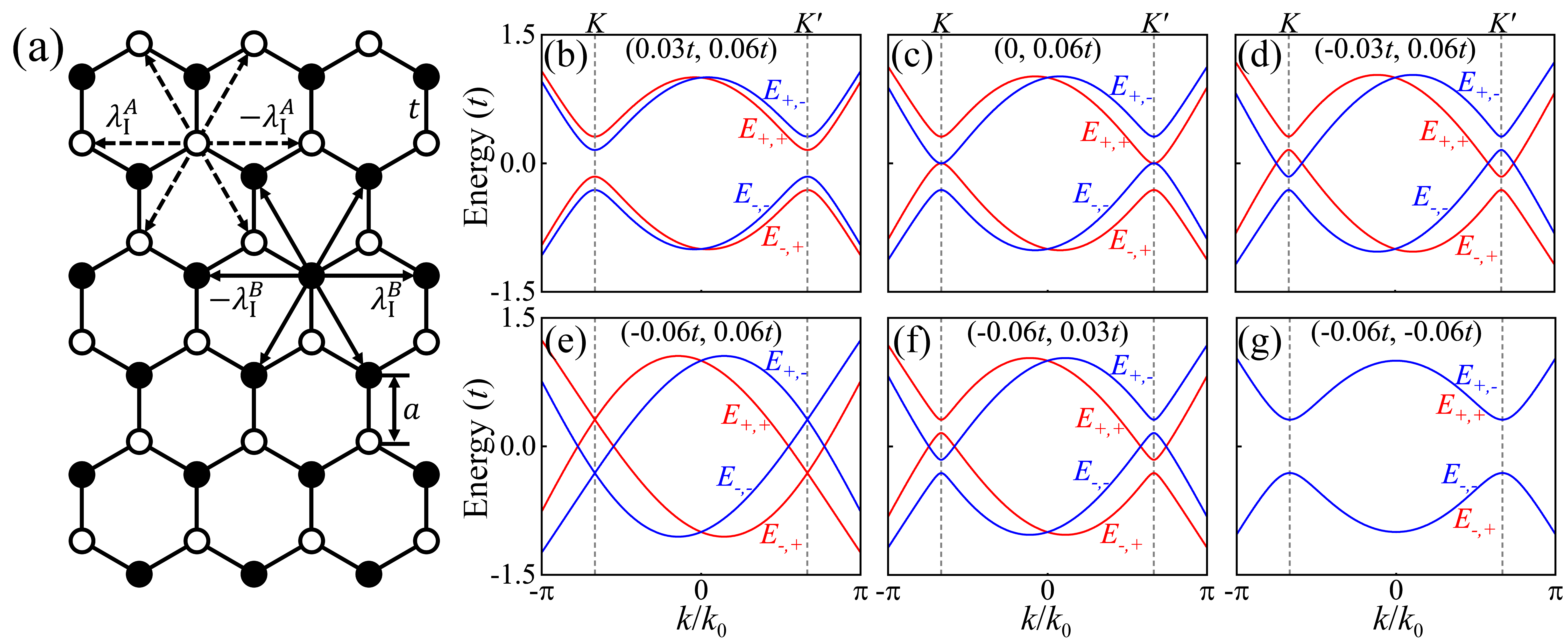}
    \caption{(a) Schematic of the MKM model. Sublattice \textit{A} is represented by empty dots, and sublattice \textit{B} by filled dots. Panels (b)-(g) show bulk energy bands along the $k_{x}$ axis at $k_{y}/k_{0}=2\sqrt{3}\pi/3$ for specific values of $\left(\lambda_{\rm I}^{A},\lambda_{\rm I}^{B}\right)$. Here, the unit of $k$ is $k_{0}=1/\sqrt{3}a$. In (f), the energy bands are double-degenerate.}
    \label{Fig2}
\end{figure*}

The original KM model encompasses several parameters that govern diverse topological quantum phases. For example, the staggered onsite energy potential $\left(\lambda_{v}\right)$ breaks the mirror symmetry in the plane, creating a trivial gap that imposes a threshold strength on the band inversion SOC strength $\left(\lambda_{\rm I}\right)$ to achieve either the QSHI phase or a trivial insulator phase \cite{Kane20051}. In this work, we set $\lambda_{v}=0$ to allow greater flexibility for $\lambda_{\rm I}$ and preserve the mirror symmetry. Regarding Rashba SOC $\left(\lambda_{\rm R}\right)$, prior research by Frank \textit{et al.} \cite{Frank2018} extensively discussed its role in realizing PHESs. In this work, we focus on AHESs, which emerge when NN Rashba SOC is absent. Therefore, we also set $\lambda_{\rm R}$ to zero for simplicity. Under these considerations, the KM model on a honeycomb lattice [Fig. \ref{Fig2}(a)] is modified to a simpler form:
\begin{equation}
    H = t\sum_{\langle{i, j}\rangle}c_{i}^{\dagger}c_{j} + i\sum_{\langle\langle{i,j}\rangle\rangle}\lambda_{\rm I}^{i}\nu_{ij}c_{i}^{\dagger}s_{z}s_{j}.
    \label{EQ1}
\end{equation}
Here, $c_{i}^{\dagger} = \left(c_{i,\uparrow}^{\dagger}, c_{i,\downarrow}^{\dagger}\right)$ is the electron creation operator at the site $i$, and $\boldsymbol{s} = \left(s_{0}, s_{x}, s_{y}, s_{z}\right)$ comprises the Pauli matrices representing electron spin. The first term on the right-hand side represents nearest-neighbor (NN) hopping with coupling strength $t$. The second term represents sublattice-dependent NNN intrinsic SOC, where $\lambda_{\rm I}^{i}$ corresponds to either $\lambda_{\rm I}^{A}$ or $\lambda_{\rm I}^{B}$ depending on whether the SOC acts on sublattice \textit{A} or \textit{B}, respectively. The parameter $\nu_{ij}=\pm 1$ corresponds clockwise or counterclockwise intra-sublattice hopping. For $\lambda_{\rm I}^{i} = \lambda_{\rm I}^{A} = \lambda_{\rm I}^{B}$, the model yields the well-known QSHI phase \cite{Kane20051}. By intentionally configuring the proximity effect \cite{Gmitra2016,Yang2016}, the KM model can be modified to host staggered intrinsic SOC, i.e. $\lambda_{\rm I}^{A}\lambda_{\rm I}^{B} < 0$ or typically $\lambda_{\rm I}^{A} = -\lambda_{\rm I}^{B}$. In these cases, helical edge states transition into either PHESs $\left(\lambda_{\rm R}\neq 0\right)$ or AHESs $\left(\lambda_{\rm R} = 0\right)$, both carrying spin-polarized currents propagate co-directionally along distinct boundaries.

Applying the Fourier transformation to Eq. (\ref{EQ1}), the Bloch Hamiltonian in \textbf{\textit{k}}-space is:
\begin{eqnarray}
    H\left(\boldsymbol{k}\right) = && f_{x}\left(\boldsymbol{k}\right)\sigma_{x}\otimes s_{0} + f_{y}\left(\boldsymbol{k}\right)\sigma_{y}\otimes s_{0} \notag\\
    && + f_{0}\left(\boldsymbol{k}\right)\left(\lambda_{\rm I}^{A}\sigma_{+} + \lambda_{\rm I}^{B}\sigma_{-}\right)\otimes s_{z},
    \label{EQ2}
\end{eqnarray}
where $\boldsymbol{\sigma}$ represents the Pauli matrices acting on sublattices, and $\sigma_{\pm}=\left(\sigma_{z}\pm\sigma_{0}\right)/2$. The coefficients are
\begin{eqnarray}
    f_{x}\left(\boldsymbol{k}\right) = t\left(1+2\cos{\frac{\sqrt{3}k_{x}a}{2}}\cos{\frac{3 k_{y}a}{2}}\right),  \notag\\
    f_{y}\left(\boldsymbol{k}\right) = -2t\cos{\frac{\sqrt{3}k_{x}a}{2}}\sin{\frac{3 k_{y}a}{2}},  \notag\\
    f_{0}\left(\boldsymbol{k}\right) = 2\sin{\sqrt{3}k_{x}a} - 4\sin{\frac{\sqrt{3}k_{x}a}{2}}\cos{\frac{3 k_{y}a}{2}}.  \notag
\end{eqnarray}
Here, $a$ is the lattice constant. The Bloch Hamiltonian $H\left(\boldsymbol{k}\right)$ is invariant under the TR operator $\hat{T} = i\sigma_{0}\otimes s_{z}\mathcal{K}$, with $\mathcal{K}$ representing complex conjugation. Note that at all the TR invariant momenta, we always have $f_{0}\left(\boldsymbol{k}\right) = 0$ in the last term of Eq. (\ref{EQ2}), implying that the values of $\lambda_{\rm I}^{i}$ will not affect the evaluation of the topological invariant $Z_{2}$ under TR protection. Diagonalizing Eq. (\ref{EQ2}) yields the energy dispersion:
\begin{eqnarray}
    E_{\gamma_{1},\gamma_{2}}\left(\boldsymbol{k}\right) && = \frac{1}{2}\gamma_{2}\left(\lambda_{\rm I}^{A}-\lambda_{\rm I}^{B}\right)f_{0}  \notag\\
    && + \frac{1}{2}\gamma_{1}\sqrt{\left(\lambda_{\rm I}^{A}+\lambda_{\rm I}^{B}\right)^{2}f_{0}^{2}+4 f_{x}^{2}+4 f_{y}^{2}}
    \label{EQ3}
\end{eqnarray}
where $\gamma_{1} = \pm 1$ is the principal index of the energy bands, and $\gamma_{2} = \pm 1$ labels the spin-up and spin-down subbands, respectively. The first term on the right-hand side emerges due to inversion symmetry breaking, which lifts the spin degeneracy and gives rise to four subbands. In Fig. \ref{Fig2}(b)-\ref{Fig2}(g), we plot the band structures for specific values of $\left(\lambda_{\rm I}^{A},\lambda_{\rm I}^{B}\right)$. These band structures suggest that quantum phase transitions rely on degeneracies at the two sets of threefold-symmetry invariant points, i.e. $K$ and $K^{\prime}$. At these points, the energies of the four subbands are given by:
\begin{equation}
    E_{\gamma_{1},\gamma_{2},\chi}=\frac{3\sqrt{3}}{2}\left[\chi\gamma_{2}\left(\lambda_{\rm I}^{A}-\lambda_{\rm I}^{B}\right)+\gamma_{1}\left|\chi\left(\lambda_{\rm I}^{A}+\lambda_{\rm I}^{B}\right)\right|\right],
    \label{EQ4}
\end{equation}
with $\chi = \pm 1$ representing the $K$ and $K^{\prime}$, respectively. Since $K$ and $K^{\prime}$ are time-reversal partner, we first focus on the $K$ point with $\chi = 1$ for simplicity. It is evident that an energy gap of size $6\sqrt{3}\lambda_{\rm I}^{i}$ always exists at $K$ and $K^{\prime}$ for $\lambda_{\rm I}^{A} = \lambda_{\rm I}^{B}$ [see Fig. \ref{Fig2}(g)]. For $\lambda_{\rm I}^{A}\neq\lambda_{\rm I}^{B}$, the two intermediate spin subbands can shift towards or away from each other depending on the relative magnitudes of $\lambda_{\rm I}^{A}$ and $\lambda_{\rm I}^{B}$. Therefore, in the parameter space $\left(\lambda_{\rm I}^{A},\lambda_{\rm I}^{B}\right)$, Eq. (\ref{EQ4}) reveals two phase boundaries related to the characteristics of band inversion at $K$: (i) For $\lambda_{\rm I}^{A} = 0$ or $\lambda_{\rm I}^{B} = 0$, the spin-polarized intermediate conduction and valence bands touch at $K$ as in Fig. \ref{Fig2}(c) when $\gamma_{1} = \chi\gamma_{2}$ or $\gamma_{1} = -\chi\gamma_{2}$. (ii) In the condition $\lambda_{\rm I}^{A} = -\lambda_{\rm I}^{B}$, the second term in Eq. (\ref{EQ4}) vanishes, the two intermediate spin subbands touch the other two spin subbands at two distinct energy points $E_{+,+,\chi}=E_{-,+,\chi}$ and $E_{+,-,\chi}=E_{-,-,\chi}$, respectively, as illustrated in Fig. \ref{Fig2}(e).

\begin{figure}
    \centering
    \includegraphics[width=3.2 in]{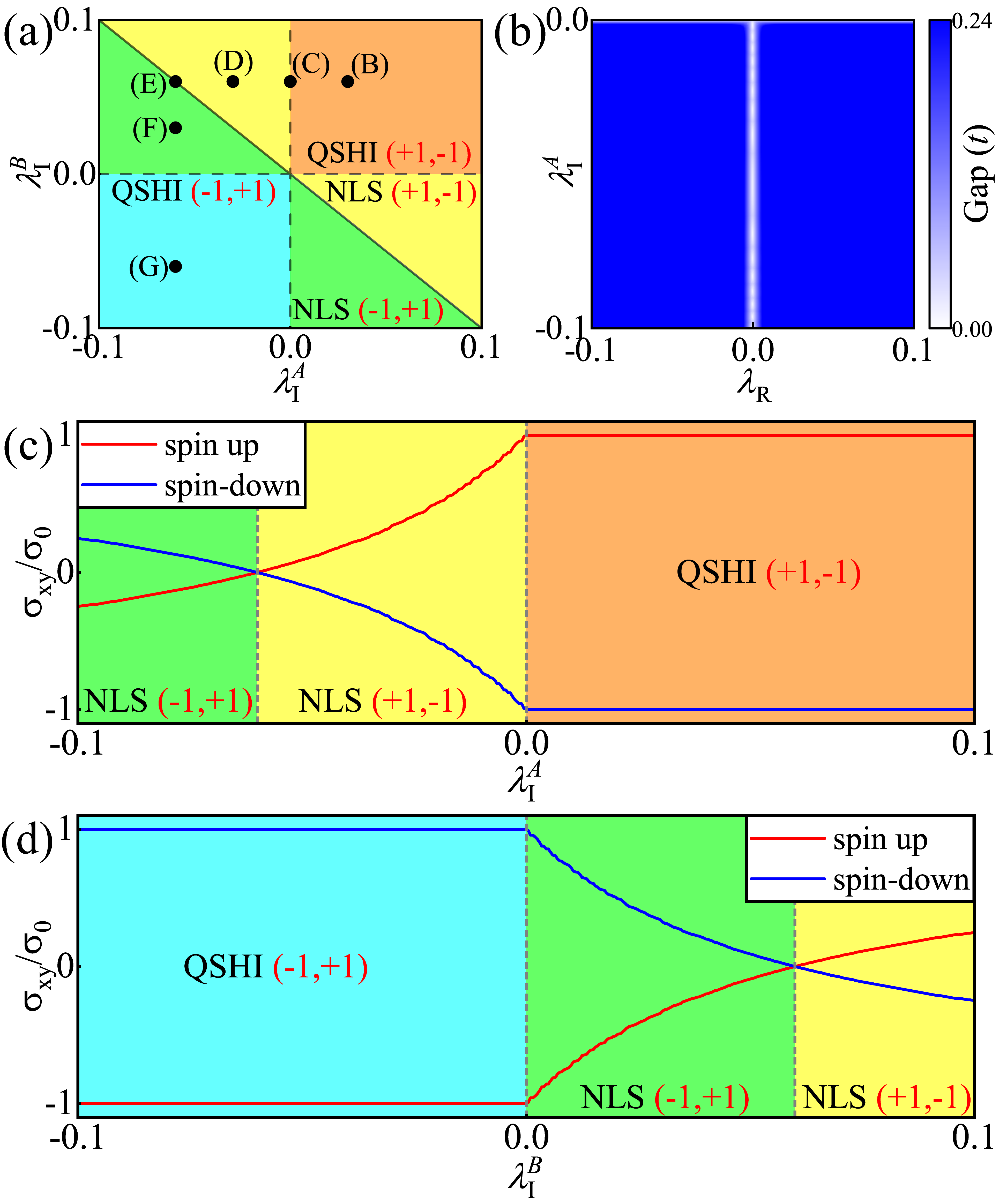}
    \caption{(a) Phase diagram of the MKM model in the parameter space of $\left(\lambda_{\rm I}^{A},\lambda_{\rm I}^{B}\right)$. The notation $\left(C_{\uparrow},C_{\downarrow}\right) = \left(\pm 1,\mp 1\right)$ represents the spin-dependent Chern number. Dashed and solid lines correspond to the phase boundaries (i) and (ii), respectively. Two QSHI phases are pained in orange and light blue, while the NLS phases are highlighted in green and yellow. (b) The map of the bulk energy band gap of the MKM model in the parameter space $\left(\lambda_{\rm R},\lambda_{\rm I}^{A}\right)$. Color denotes the size of the gap, and the ratio $\lambda_{\rm I}^{A}/\lambda_{\rm I}^{B}$ is fixed at $-1$. (c) and (d) The spin-dependent Hall conductance in units of $\sigma_{0}=e^{2}/h$ as a function of (c) $\lambda_{\rm I}^{A}$ and (d) $\lambda_{\rm I}^{B}$ at $E_{F} = 0$. In (c), $\lambda_{\rm I}^{B}$ is set to $0.06t$, while $\lambda_{\rm I}^{A}$ for (d) is fixed at $-0.06t$. The colors corresponding to different phases are the same as the phase diagram in (a). The thermal energy $k_{B}T$ is set to $10^{-6}$.}
    \label{Fig3}
\end{figure}

Using two sets of phase boundaries, we construct the phase diagram in the parameter space of $\left(\lambda_{\rm I}^{A},\lambda_{\rm I}^{B}\right)$ in Fig. \ref{Fig3}(a). The phases marked in the second quadrant are exactly equivalent to those in the fourth quadrant, since they are symmetric under the interchange of $\lambda_{\rm I}^{A}$ and $\lambda_{\rm I}^{B}$ in the Hamiltonian Eq. (\ref{EQ1}). The bulk band characteristics of each phase are illustrated in Figs. \ref{Fig2}(b)-\ref{Fig2}(g), corresponding to the points labelling as (B)-(G). In the orange and light blue zones of Fig. \ref{Fig3}(a), $\lambda_{\rm I}^{A}$ and $\lambda_{\rm I}^{B}$ always hold the same sign $\left(\lambda_{\rm I}^{A}\lambda_{\rm I}^{B}>0\right)$ and the staggered SOC effect is weak, so that the intrinsic SOC maintains the role to fully gap the bulk bands, as seen in Figs. \ref{Fig2}(b) and \ref{Fig2}(g). Therefore, we identified these two zones with the QSHI phases. As the parameters $\left(\lambda_{\rm I}^{A},\lambda_{\rm I}^{B}\right)$ vary along the path from (B) to (D), the bulk gap closes when arriving at point (C) on the boundary $\lambda_{\rm I}^{A} = 0$ and the system transitions into a bulk metallic phase. As the parameters continue to deviate from (C) to (D) in the yellow zone, band inversion occurs, and the system becomes the Weyl NLS phase, featured by the emergence of the nodal lines round the $K$ and $K^{\prime}$. Note that the system retains mirror symmetry $\left(M_{z}\right)$. This scenario is very similar to the Dirac line nodes in inversion-symmetric crystals \cite{Kim2015} and other Weyl NLS phase \cite{Liu2018}, except that we now start with a QSHI instead of a trivial insulator, and a Weyl nodal line of two-fold degeneracy emerges and grows. Applying the $Z_{2}$ formalism developed for NLS \cite{Lu2017}, we find that the present Weyl NLS possesses a nontrivial $Z_{2}$ topological invariant of $1$, which dictates the presence of an odd number of Weyl nodal lines in half the Brillouin zone. Therefore, the present Weyl NLS also represents a new $Z_{2}$ class of topological metal with the presence of staggered intrinsic SOC.

In Fig. \ref{Fig3}(a), the phase boundary $\lambda_{\rm I}^{A}=-\lambda_{\rm I}^{B}$ divides the phase diagram into two off-diagonal blocks that differ only in the signs of spin-dependent Hall conductance. Following Sheng’s scheme \cite{Sheng2006}, we employ the spin-dependent Chern numbers $\left(C_{\uparrow},C_{\downarrow}\right)$ to mark the details of bulk topology. The spin-dependent Hall conductance can be numerically calculated with knowledge of the bulk states. In Fig. \ref{Fig3}(c) and \ref{Fig3}(d), the spin-dependent Hall conductance is plotted as a function of (c) $\lambda_{\rm I}^{A}$ and (d) $\lambda_{\rm I}^{B}$, respectively. The signs are exclusively related to the spin-dependent Chern numbers $\left(C_{\uparrow},C_{\downarrow}\right)$. The result of $Z_{2}=\frac{C_{\uparrow}-C_{\downarrow}}{2}\;{\rm mod}\; 2=1$ \cite{Hasan2010} further confirms the non-vanishing $Z_{2}$ invariant in both QSHI and NLS phases. In the QSHI phases, the helical edge states contribute to the quantized spin Hall conductance $\sigma_{xy}^{s}=\sigma_{xy}^{\uparrow}-\sigma_{xy}^{\downarrow}=\pm 2 e^{2}/h$ \cite{Kane20051}. While for the NLS phases, spin Hall conductance is not quantized due to the existence of the gapless bulk states \cite{Ying2018,Ying2019,Lee20241,Lee20242}. Moreover, from Fig. \ref{Fig3}(c) and \ref{Fig3}(d), one can learn that the spin Hall conductance reverse signs when crossing the boundary $\lambda_{\rm I}^{A}=-\lambda_{\rm I}^{B}$. Therefore, from the viewpoint of spin-dependent Chern number and Hall conductance, the phase diagram Fig. \ref{Fig3}(a) is antisymmetric about the boundary $\lambda_{\rm I}^{A}=-\lambda_{\rm I}^{B}$.

At the end of this section, we briefly comment on the effect of Rashba SOC on the bulk topological properties. Note that the Rashba SOC does not change the QSHI phase as long as $\lambda_{\rm R}$ is not too large \cite{Dai2024,Kane20052}, but it does play an important role in triggering phase transitions from the NLS phase to an insulator phase. In Fig. \ref{Fig3}(b), we illustrate that an energy gap opens at the anterior NLS phase $\left(\lambda_{\rm I}^{A}=-\lambda_{\rm I}^{B}\right)$ in the presence of Rashba SOC. When the Rashba SOC is included, the nodal lines are gapped out, and the system transitions into the $Z_{2}$-trivial insulating phase \cite{Frank2018}. In this circumstance, the Rashba SOC induces spin flips along the armchair boundaries, connecting the co-directionally propagating spin-polarized edge modes at opposite zigzag-terminated boundary \cite{Frank2018,Zhumagulov2022}. This differs from the AHESs, in which \textit{no} spin flip is observed between the edge states and their compensating bulk states, as will be discussed in next section.

\section{\label{sec:Edge States}Edge States}

\begin{figure}
    \centering
    \includegraphics[width=3.4 in]{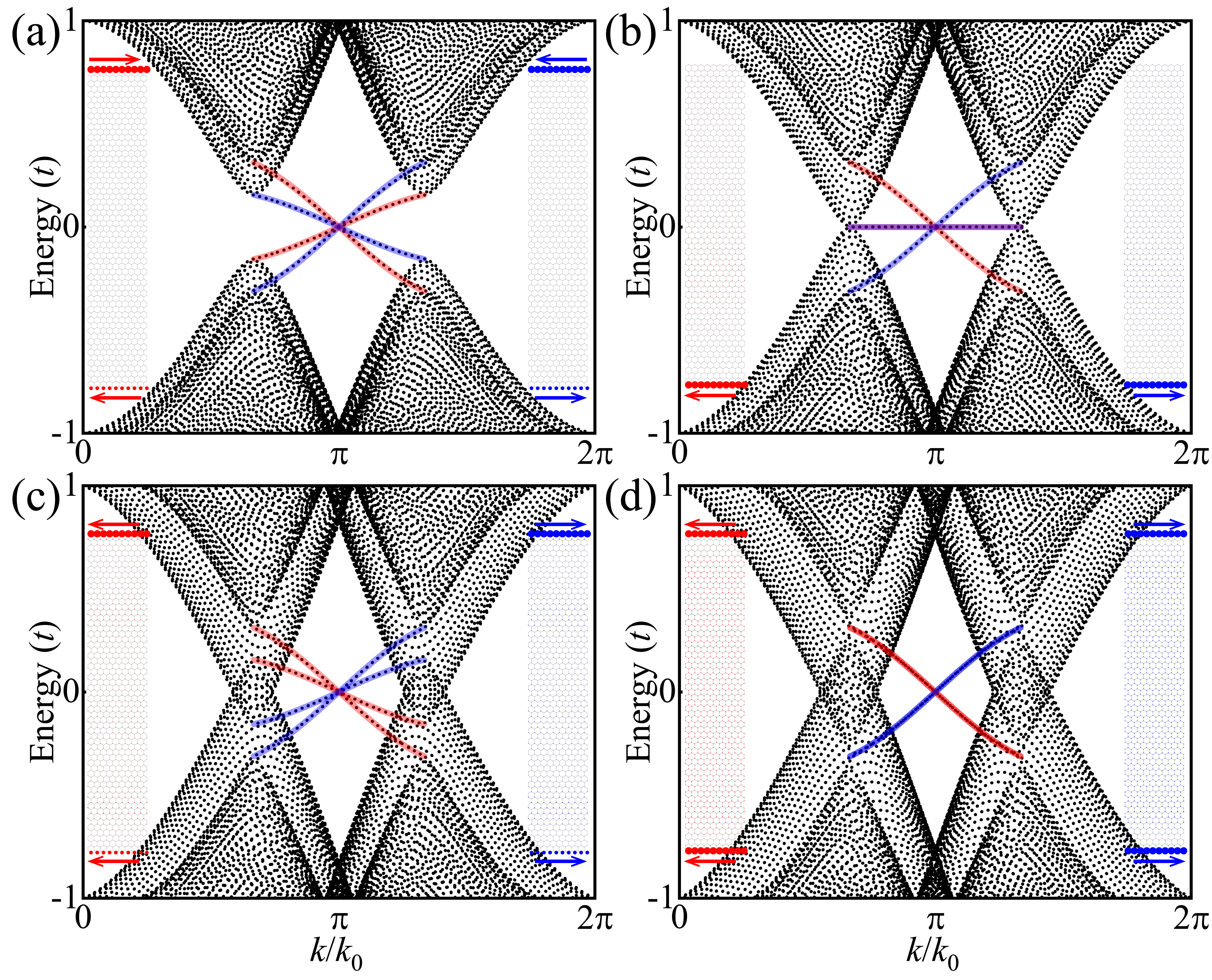}
    \caption{The energy spectra of a ZNR and corresponding spatial local density of states for spin-up (left inset) and spin-down (right inset) electrons at $E_{F}=0.03t$ for various $\left(\lambda_{\rm I}^{A},\lambda_{\rm I}^{B}\right)$. The dotted lines represent the numerical energy spectra of a ZNR with width $W=89a$. The analytical representations of the edge energy dispersion of the edge states, obtained from Eqs. (\ref{EQ5}) and (\ref{EQ6}), are depicted by red (for spin-up) and blue (for spin-down) heavy lines. The parameters $\left(\lambda_{\rm I}^{A},\lambda_{\rm I}^{B}\right)$ in (a)-(d) correspond to the coordinates of points (B)-(E) in the phase diagram in Fig. \ref{Fig3}(a).}
    \label{Fig4}
\end{figure}

There are generally two nanoribbon geometries in honeycomb materials: the zigzag edge termination or armchair edge termination. In this work, we focus on the zigzag-terminated nanoribbon, denoted as zigzag nanoribbon or ZNR. For the armchair nanoribbon, it generally introduces the inter-valley coupling that could break the edge states, especially in the presence of valley Zeeman term in certain MKM models \cite{Frank2018}. In Fig. \ref{Fig4}, we plot the complete spectra for ZNRs, with parameters $\left(\lambda_{\rm I}^{A},\lambda_{\rm I}^{B}\right)$ corresponding to the phase points (B) to (E) in Fig. \ref{Fig3}(a). The fully gapped bulk bands of QSHI phase and four in-gap helical edge states are well visible in Fig. \ref{Fig4}(a). When the parameters $\left(\lambda_{\rm I}^{A},\lambda_{\rm I}^{B}\right)$ reach (C) point, two of the four helical edge states evolve into a double-degenerate flat band, akin to that of a zigzag-terminated graphene nanoribbon, as depicted in Fig. \ref{Fig4}(b). In the staggered cases, as seen in Fig. \ref{Fig4}(c) and \ref{Fig4}(d), the band inversions at $K$ and $K^{\prime}$ cause these two flat bands to disperse again. However, compared with the helical cases, they flip their propagation direction, causing edge states to transform into the antihelical type. In particular, at $\lambda_{\rm I}^{A}=-\lambda_{\rm I}^{B}$, the edge states with same spin at opposite boundaries host the same energy dispersion relation. The insets in each panel are the spatial local density of states for spin-up (on the left of the spectra) and spin-down (on the right) electrons, respectively. As shown in the insets, the local density of states for both spins is most concentrated along the sample boundaries.

In the absence of the staggered onsite potential and the Rashba SOC, we do not expect the additional valley edge states for the 2D NLS. Generally, the 2D NLSs do not possess protected edge states enclosed by the projection of the nodal line as in a 3D NLS, because the reduced sample dimension leads to the vanishing co-dimension of a nodal lines \cite{Jin2017,Feng2021}. Therefore, there are only one pair of AHESs at each boundary, which has no connection with the nodal lines. We should emphasize that these AHESs are intrinsically different from the PHESs. The AHESs can be viewed as two spin copies of antichiral edge modes, and the gapless bulk states resided in the same energy window provide compensation channels for the imbalanced edge spin currents. Similar to the antichiral edge modes in the modified Haldane model, the edge and the bulk modes must also be spatially separated. In this sense, the backscattering of the edge modes could be strongly suppressed \cite{Colomes2018,Mandal2019}, provided that the coupling of edge modes and bulk modes are neglected \cite{mannai2023}.

To gain further insight into these spin-polarized edge states, we work out the approximate analytical expression with a perturbative treatment following the approach outlined in Ref. \cite{Rahmati2023}. The energy dispersion relations of the edge states at $\lambda_{\rm I}^{A}\lambda_{\rm I}^{B}\ll t^{2}$ can be approximately cast into two expressions [see Appendix \ref{sec:appendixA} for details]
\begin{equation}
    E_{s}^{l}\left(k\right)=6s\lambda_{\rm I}^{B}\sin{\sqrt{3}ka}
    \label{EQ5}
\end{equation}
for lower boundary, and
\begin{equation}
    E_{s}^{u}\left(k\right)=-6s\lambda_{\rm I}^{A}\sin{\sqrt{3}ka}
    \label{EQ6}
\end{equation}
for upper boundary. Here, $s=\pm 1$ represents the spin-up and spin-down edge modes, respectively. Eqs. (\ref{EQ5}) and (\ref{EQ6}) are plotted in Fig. \ref{Fig4}, with the red (for spin-up) and blue (for spin-down) heavy lines showing good agreement with the dotted numerical results. From Eqs. (\ref{EQ5}) and (\ref{EQ6}), one can immediately see that (i) the energy dispersion relations of the edge states depend solely on the strength of intrinsic SOCs acting on sites \textit{A} and \textit{B}; (ii) each of the \textit{A} and \textit{B} sublattices contributes two spin polarized edge states to each boundary; (iii) for $\lambda_{\rm I}^{A}=-\lambda_{\rm I}^{B}$, the edge states manifest exactly the same at opposite boundaries. The latter implies a crucial condition for realizing the PHESs and the AHESs.

Additionally, all the spin polarized edge states obey the spin-momentum locking under the protection of TR symmetry. One can inspect on the effective velocities: $v_{s}^{l}\left(k\right)=s\frac{6\sqrt{3}a\lambda_{\rm I}^{B}\cos{\sqrt{3}ka}}{\hbar}$ and $v_{s}^{u}\left(k\right)=-s\frac{6\sqrt{3}a\lambda_{\rm I}^{A}\cos{\sqrt{3}ka}}{\hbar}$. For any given intrinsic SOCs $\left(\lambda_{\rm I}^{A},\lambda_{\rm I}^{B}\right)$, the direction of the velocities relies solely on the value of the spin index $s$. For example, if the spin index $s$ takes the same value for both lower and upper boundaries, one expects to observe the helical edge states for $\lambda_{\rm I}^{A}\lambda_{\rm I}^{B}>0$ (the QSHI phases), and the AHESs for $\lambda_{\rm I}^{A}\lambda_{\rm I}^{B}<0$ (the 2D NLS phases). Typically, when $\lambda_{\rm I}^{A}=-\lambda_{\rm I}^{B}$, the AHESs at opposite boundaries are spatially mirror symmetry.

\begin{figure}
    \centering
    \includegraphics[width=3 in]{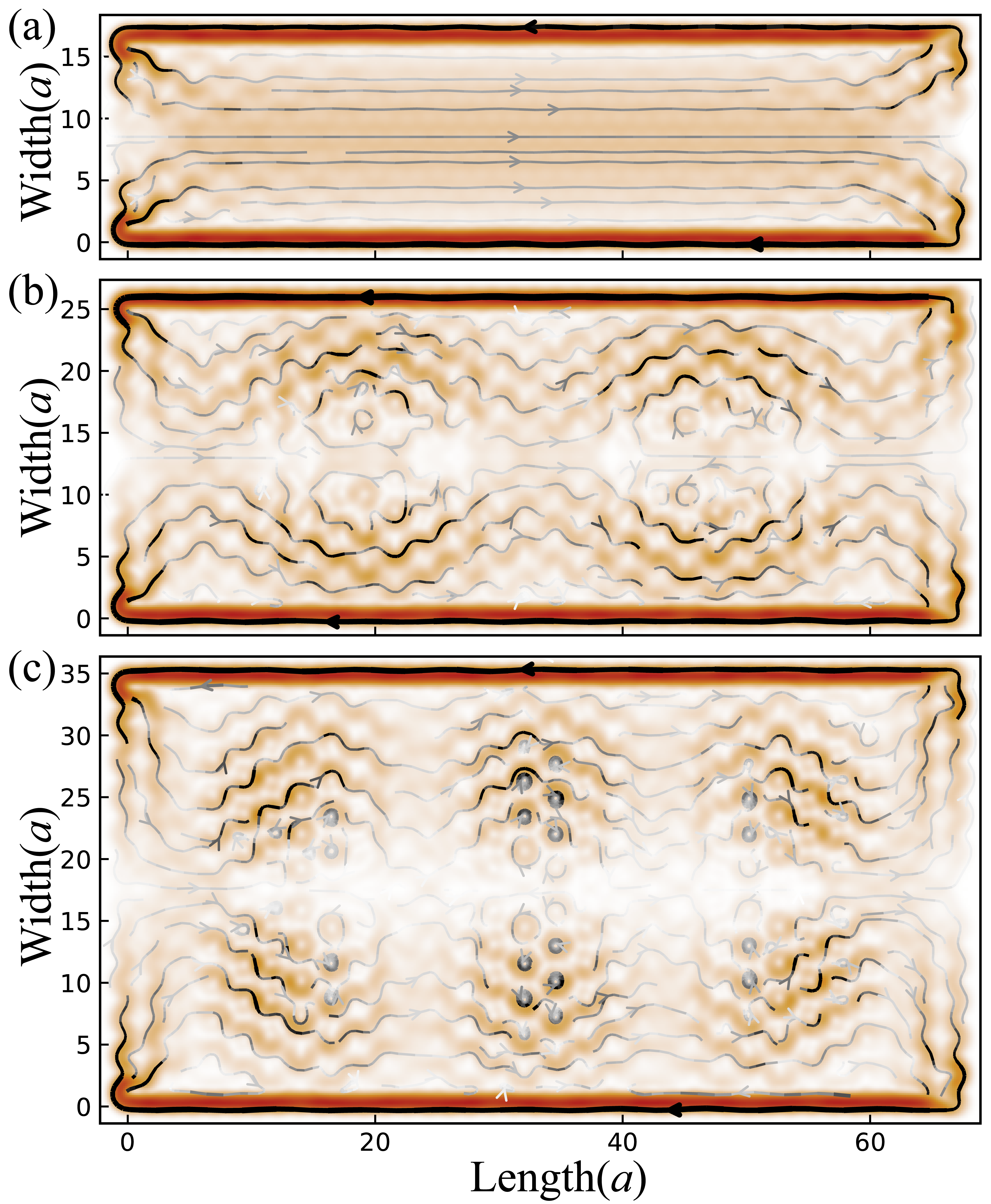}
    \caption{The spin current of the highest occupied state for $\lambda_{\rm I}^{A}=-\lambda_{\rm I}^{B}$. The size of the flake is chosen as (a) $79\sqrt{3}a/2\times 17a$, (b) $79\sqrt{3}a/2\times 26a$ and (c) $79\sqrt{3}a/2\times 35a$, which correspond to $79\times 23$, $79\times 35$ and $79\times 74$ NN bond sized flake, respectively.}
    \label{Fig5}
\end{figure}

To see how the AHESs meet in a finite flake, we show in Fig. \ref{Fig5} the spin current of the highest occupied states which corresponds to the Fermi energy of $E_{F}\approx 0$ for different flake widths. The absence of Rashba SOC means that spin flip cannot occur at armchair boundaries, making “perfect tunneling” between opposite zigzag boundaries impossible. Moreover, due to the TR symmetry, the backscattering between the edge states at the same zigzag boundary is also strictly forbidden. Therefore, at the armchair boundaries, the AHESs can only reflect back into the bulk states. This is similar to the perfect reflection mechanism that has been described to explain the stability of the PHESs \cite{Zhumagulov2022} in the presence of a magnetic field. In the present scenario, for any flake size, the states close to the Fermi energy must be a mixture of strongly localized AHESs and delocalized bulk states. The number of bulk states depends on the size of the flake. For a narrow flake of width $17a$, the spin current pattern follows the perfect reflection mechanism, as illustrated in Fig. \ref{Fig5}(a). As the width increases, more bulk states emerge near the Fermi energy. In the bulk limit, the bulk spin currents not only reflect into the edge, but also reflect back to the bulk and interfere with each other to form the standing wave patterns, as depicted in Fig. \ref{Fig5}(b) and \ref{Fig5}(c). Therefore, the AHESs should be distinguished from those PHESs in finite-size systems.

\section{\label{sec:Robustness of Edge States}Robustness of the Edge States}

\begin{figure}
    \centering
    \includegraphics[width=3.4 in]{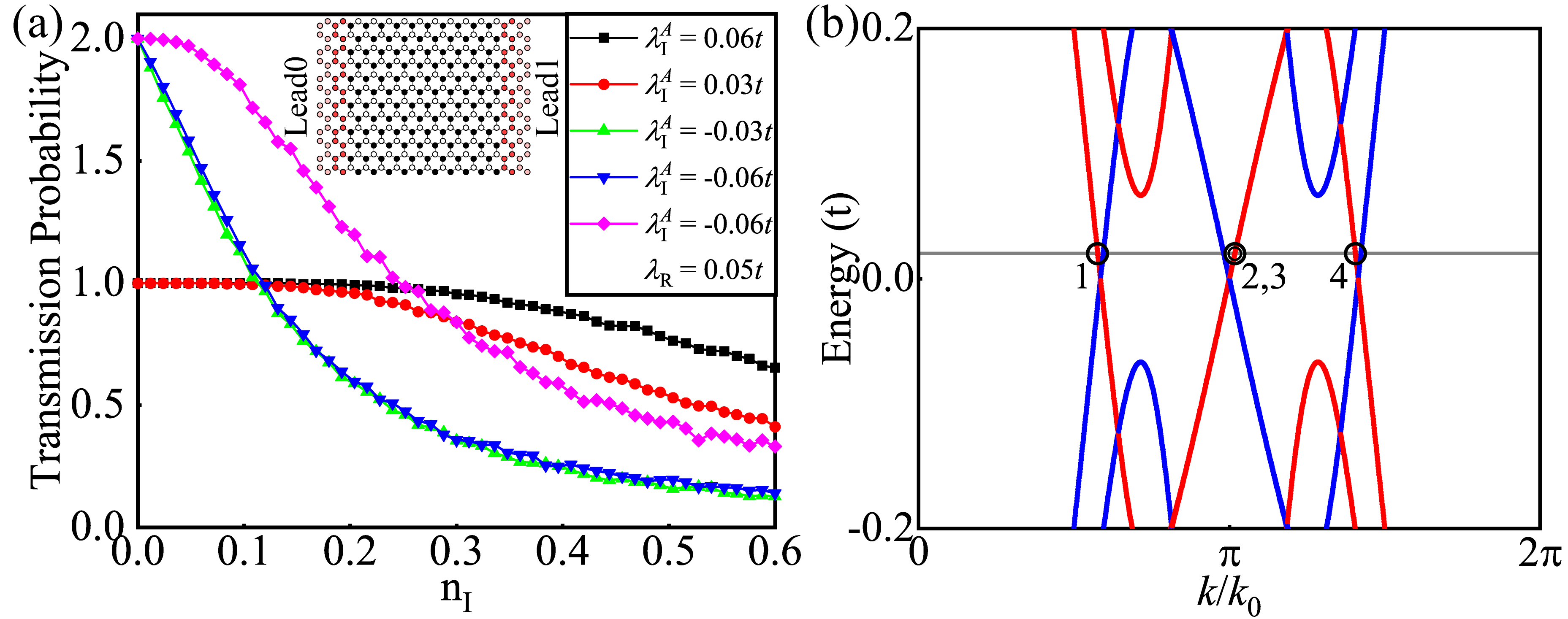}
    \caption{(a) Transmission probability of spin-up electron in an MKM sample with size $79\sqrt{3}a/2\times 17a$ from Lead0 to Lead1 (see scheme shown in the inset) as a function of the disorder concentration ${\rm n}_{\rm I}$ for $\lambda_{\rm I}^{B}=0.06t$, $U=5t$, and $E_{F}=0.01t$. Different $\lambda_{\rm I}^{A}$ values are depicted. The magenta line represents the transmission probability contributed by PHESs with Rashba SOC strength $\lambda_{\rm R}=0.05t$.  (b) Band structure of a ZNR with width $17a$ and hopping parameters $\lambda_{\rm I}^{A}=-\lambda_{\rm I}^{B}=-0.06t$. Color encodes spin-expectation, where red and blue stand for spin-up and spin-down, respectively. Two spin-up double-degenerate antihelical edge channels (2 and 3) and two bulk channels (1 and 4) are involved in the transmission process.}
    \label{Fig6}
\end{figure}

All the aforementioned TR invariant topological edge states have been predicted to be robust against nonmagnetic impurities. To evaluate the robustness of the topological edge states, we compute the multi-channel transmission coefficient for three types of edge states of the MKM model using the Kwant package \cite{Groth2014}. The total transmission coefficients are obtained by summing over the contributions from all the channels involved in the transmission process \cite{Mella2022}. The device scheme for simulation is shown in the inset of Fig. \ref{Fig6}(a). Here, the size of MKM sample is $79\sqrt{3}a/2\times 17a$, and the source (Lead0) and collector (Lead1) are fully contacted with the left and right armchair boundaries. We expect that the bulk modes do not contribute to the total transmission coefficient in the 2D NLS sample, for the fact that the edge and bulk spin modes transport in opposite velocity. The nonmagnetic onsite disorder is simulated by the term $H_{\rm dis}=\sum_{i}V_{\rm d}\left(i\right)c_{i}^{\dagger}c_{i}$, where $V_{\rm d}\left(i\right)=Uw_{i}$ with $U$ representing the disorder strength and $w_{i}\in\left[-1/2,1/2\right]$ being uniformly distributed random numbers \cite{Lu2023}. To eliminate sample fluctuations, all transmission results are averaged over $1000$ identical systems with different initial seeds in the random disorders. For simplicity, the leads are considered to share the same tight-binding Hamiltonian as the samples.

Figure \ref{Fig6}(a) displays the total transmission probability of spin-up electron against the disorder concentration, ${\rm n}_{\rm I}$. Note that the odd TR symmetry ensures identical transmission characteristics for spin-down cases. The initial values correspond to ballistic devices, where we find that the helical edge states contribute one mode per two boundaries due to their helical nature, and both the PHESs and the AHESs contribute one mode per each boundary. Noted that the valley edge states have been gapped out by the size quantization of the narrow sample, and no bulk modes contribute to the antihelical edge channels, as expected.

As shown by the black and red curves in Fig. \ref{Fig6}(a), the transmission probability of the helical edge states manifests strong robustness against the nonmagnetic disorder up to a concentration of about ${\rm n}_{\rm I}=0.21N$ for $\lambda_{\rm I}^{A}=0.06t$ and ${\rm n}_{\rm I}=0.12N$ for $\lambda_{\rm I}^{A}=0.03t$, with $N$ being the total number of sites in the system. At higher disorder concentrations, the inter-edge scattering occurs, wherein an electron traveling along the upper boundary tunnels into the lower edge states through multiple disorder-induced scattering processes \cite{Vannucci2020}. For the PHESs [magenta curve in Fig. \ref{Fig6}(a)], the transmission probability tends to mimic that of the helical edge states but with a very low threshold disorder concentration of about ${\rm n}_{\rm I}=0.04N$. Due to the nature of co-directional propagation of PHESs, the inter-edge scattering between the edge states with the same spin does not lead to a decrease in transmission probability. We account this decrease originates from the inter-edge scattering between edge states with the opposite spins, in which the spin flips are realized by Rashba SOC, similar to the situation when PHESs meet at the armchair boundaries. Therefore, the insulting bulk not only acts as a potential barrier but also flips the spin of electrons. Moreover, due to the smaller gap determined by Rashba SOC, inter-edge scattering is more likely to occur among the PHESs than the helical ones, indicating that PHESs are less robust than the helical edge states \cite{Frank2018}.

For AHESs, the transmission probabilities exponentially decay with increasing disorder concentration and break down at high concentrations, as depicted by blue and green curves in Fig. \ref{Fig6}(a). The fragility of AHESs is pretty similar to that of antichiral edge states \cite{Yu2021,Bao2022}, in which the antichiral edge states can scatter into bulk states resided in the same energy window in the presence of edge disorders. Although some works predicted that antichiral/antihelical edge states are robust against disorder due to the spatial separation between the edge states and bulk states \cite{Colomes2018,Mandal2019,Xie2023}, the scattering between edge states and bulk states can become significant in the presence of edge disorder, which is possible given the randomly distributed disorders considered. This fragility has also been demonstrated in the modified Haldane model by Manna{\"\i} \textit{et al.} \cite{mannai2023}, in which the topological invariant is suppressed, and the antichiral edge states are easily localized by small disorder amplitude. Therefore, the suppression of quantized transmission probability of AHESs is a consequence of the scattering between the AHESs and bulk states [i.e., edge states 2, 3 and bulk states 1, 4 in Fig. \ref{Fig6}(b)], akin to those of antichiral edge states \cite{mannai2023,Yu2021,Bao2022}

Therefore, the AHESs protected by the TR symmetry are significantly more fragile against disorders than PHESs and helical edge states. However, we view this as a potential opportunity to distinguish these three anomalous topological edge states, and suggest that AHESs as a necessary ingredient to the MKM model.

\section{\label{sec:Conclusion and Discussion}Conclusion and Discussion}

In conclusion, we explored the topological properties of a MKM model by making the NNN intrinsic SOC strengths of the two sublattices unequal, thus breaking the inversion symmetry of the Hamiltonian. This inversion symmetry breaking alters the band gap, leading the MKM model to undergo a phase transition at $\lambda_{\rm I}^{A}\lambda_{\rm I}^{B}=0$, shifting from the QSHI phases with quantized spin Hall conductance to 2D Weyl NLS phases with unquantized spin Hall conductance. These phases are characterized by a $Z_{2}$ topological invariant and distinguished by opposite spin-dependent Chern numbers. In a strip configuration with zigzag termination, we observed that as the intrinsic SOCs drive the MKM model into NLS from the QSHI, the edge states strongly localized at one boundary reverse their propagation direction. This transition results in the transformation of helical edge states into AHESs which can be viewed as a superposition of two spin copies of antichiral edge states. The gapless bulk nature of NLS phases provides compensating channels for the AHESs. When these AHESs encounter armchair boundaries, they perfectly reflect into the bulk states, a behavior confirmed by the spin current in a finite-size flake. Finally, we addressed the robustness of the AHESs in the MKM model using a scattering perspective. Unlike helical and PHESs, in which multiple disorders open the scattering channel between the edge states at opposite boundaries, causing these edge states are robust against low concentration disorders, the AHESs are much fragile because the edge-bulk scattering induced by edge disorders occurs more readily than inter-edge scattering.

We also notice that the AHESs can also be realized by breaking the inversion symmetry of a 2D square lattice through tuning nearest-neighbor (NN) coupling, resulting in AHESs in a topological metal phase \cite{Xie2023}. 
Additionally, another approach to achieve AHESs was proposed by Li \cite{Li2023}, in which two antiferromagnetic triangular lattices with opposite Dzyaloshinskii-Moriya interactions are combined head-by-head into a heterostructure. 
In this system, each subsystem contributes a set of copropagating spin currents on opposite boundaries to form AHESs.

Experimentally, staggered intrinsic SOC can be introduced by placing graphene on transition metal dichalcogenides \cite{Wang2015,Gmitra2016,Yang2016,Wang2016,Avsar2020,Khatibi2022}. However, in these systems, the $M_{z}$ symmetry is absent, thus allowing non-vanishing Rashba SOC and only promising the PHESs. Nonetheless, an LC circuit network offers a promising avenue for achieving NLS phases and AHESs, in which the staggered intrinsic SOC can be controlled by the manipulation of capacitances and inductances \cite{Zhu2019}. Therefore, the AHESs proposed in this work hold potential application in spin-polarized signal transportation.

\begin{acknowledgments}
X. Dai would like to thank Y. Yang and H. Zhao for fruitful discussions. This work was supported by the Natural Science Foundation of Hunan Province, China (Grant No. 2023JJ30114). P.-H. F. and Y. S. A. are supported by the Singapore Ministry of Education (MOE) Academic Research Fund (AcRF) Tier 2 Grant (MOE-T2EP50221-0019).
\end{acknowledgments}

\appendix
\section{\label{sec:appendixA}Derivation of Energy Dispersion of Edge States}
\begin{figure}
    \centering
    \includegraphics[width=3 in]{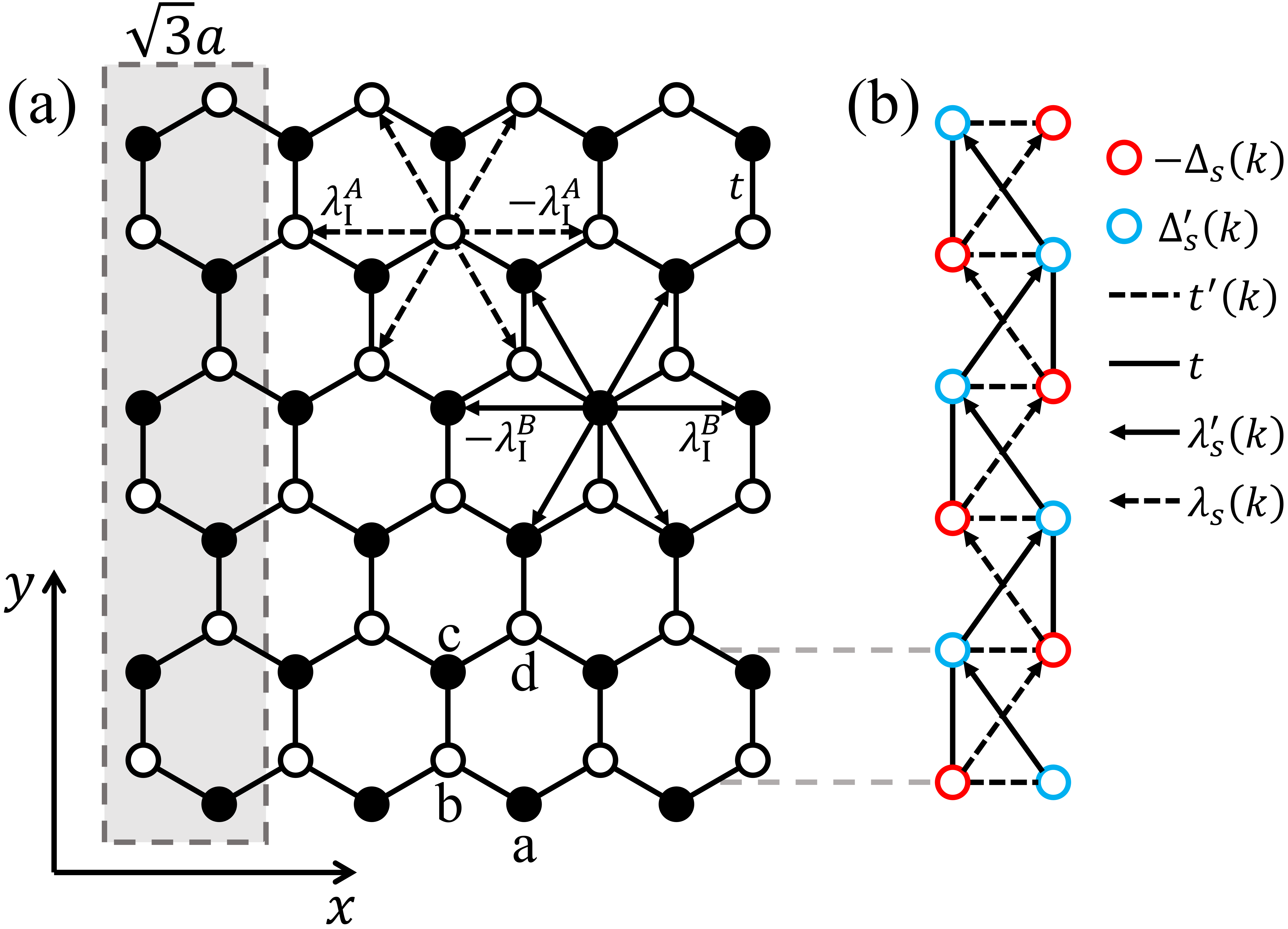}
    \caption{(a) Schematic of a one-dimensional zigzag-terminated MKM nanoribbon and (b) corresponding two-leg ladder system.}
    \label{Fig7}
\end{figure}

In the zigzag-terminated MKM model, which is periodic in the $x$ direction and finite along the $y$ direction, the system can be equivalently represented as a two-leg ladder model with momentum-dependent hopping parameters after applying a Fourier transformation to $x$ direction \cite{Rahmati2023}, as depicted in Fig. \ref{Fig7}. The Hamiltonian of this two-leg ladder model can be divided into two parts: the unperturbed part,
\begin{equation}
    \begin{aligned}
        H_{0}\left(k\right) = && \sum_{m}\Big\{ \phi_{k,m}^{\dagger}T_{0}\left(k\right)\phi_{k,m}  
        +\phi_{k,m+1}^{\dagger}T_{1}\left(k\right)\phi_{k,m} \\ &&+\phi_{k,m}^{\dagger}T_{1}^{\dagger}\left(k\right)\phi_{k,m+1}\Big\},     
    \end{aligned}
    \label{EQA1}
\end{equation}
and the perturbed part
\begin{equation}    
    H_{1}\left(k\right)=\sum_{m}\phi_{k,m}^{\dagger}T_{2}\left(k\right)\phi_{k,m},
\label{EQA2}
\end{equation}
where $\phi_{k,m}^{\dagger}=\left({\rm a}_{k,m}^{\dagger},{\rm b}_{k,m}^{\dagger},{\rm c}_{k,m}^{\dagger},{\rm d}_{k,m}^{\dagger}\right)$ with ${\rm a}$, ${\rm b}$, ${\rm c}$ and ${\rm d}$ representing the sublattices in the plaquette [see Fig. \ref{Fig7}(a)]. For brevity, the spin index is omitted. The matrices $T_{0}\left(k\right)$, $T_{1}\left(k\right)$ and $T_{2}\left(k\right)$ in Eqs. (\ref{EQA1}) and (\ref{EQA2}) are defined as
\begin{equation}
    T_{0}\left(k\right)=\left[
    \begin{array}{cccc}
        0 & 0 & -\lambda^{\prime}\left(k\right) & 0  \\
        0 & 0 & t & \lambda\left(k\right) \\
        -\lambda^{\prime}\left(k\right) & t & 0 & 0 \\
        0 & \lambda\left(k\right) & 0 & 0
    \end{array}
    \right],  \notag
\end{equation}
\begin{equation}
    T_{1}\left(k\right)=\left[
    \begin{array}{cccc}
        0 & 0 & -\lambda^{\prime}\left(k\right) & t \\
        0 & 0 & 0 & \lambda\left(k\right) \\
        0 & 0 & 0 & 0 \\
        0 & 0 & 0 & 0
    \end{array}
    \right],  \notag
\end{equation}
\begin{equation}
    T_{2}\left(k\right)=\left[
    \begin{array}{cccc}
        \Delta^{\prime}\left(k\right) & t^{\prime}\left(k\right) & 0 & 0 \\
        t^{\prime}\left(k\right) & -\Delta\left(k\right) & 0 & 0 \\
        0 & 0 & \Delta^{\prime}\left(k\right) & t^{\prime}\left(k\right) \\
        0 & 0 & t^{\prime}\left(k\right) & -\Delta\left(k\right)
    \end{array}
    \right],  \notag
\end{equation}
with momentum-dependent hopping amplitudes: $t^{\prime}\left(k\right)=2t\cos{\frac{\sqrt{3}k a}{2}}$, $\lambda\left(k\right)=2s\lambda_{\rm I}^{A}\sin{\frac{\sqrt{3}k a}{2}}$, $\lambda^{\prime}\left(k\right)=2s\lambda_{\rm I}^{B}\sin{\frac{\sqrt{3}k a}{2}}$, $\Delta\left(k\right)=2s\lambda_{\rm I}^{A}\sin{\sqrt{3}k a}$, and $\Delta^{\prime}\left(k\right)=2s\lambda_{\rm I}^{B}\sin{\sqrt{3}k a}$.

\begin{figure}
    \centering
    \includegraphics[width=3.4 in]{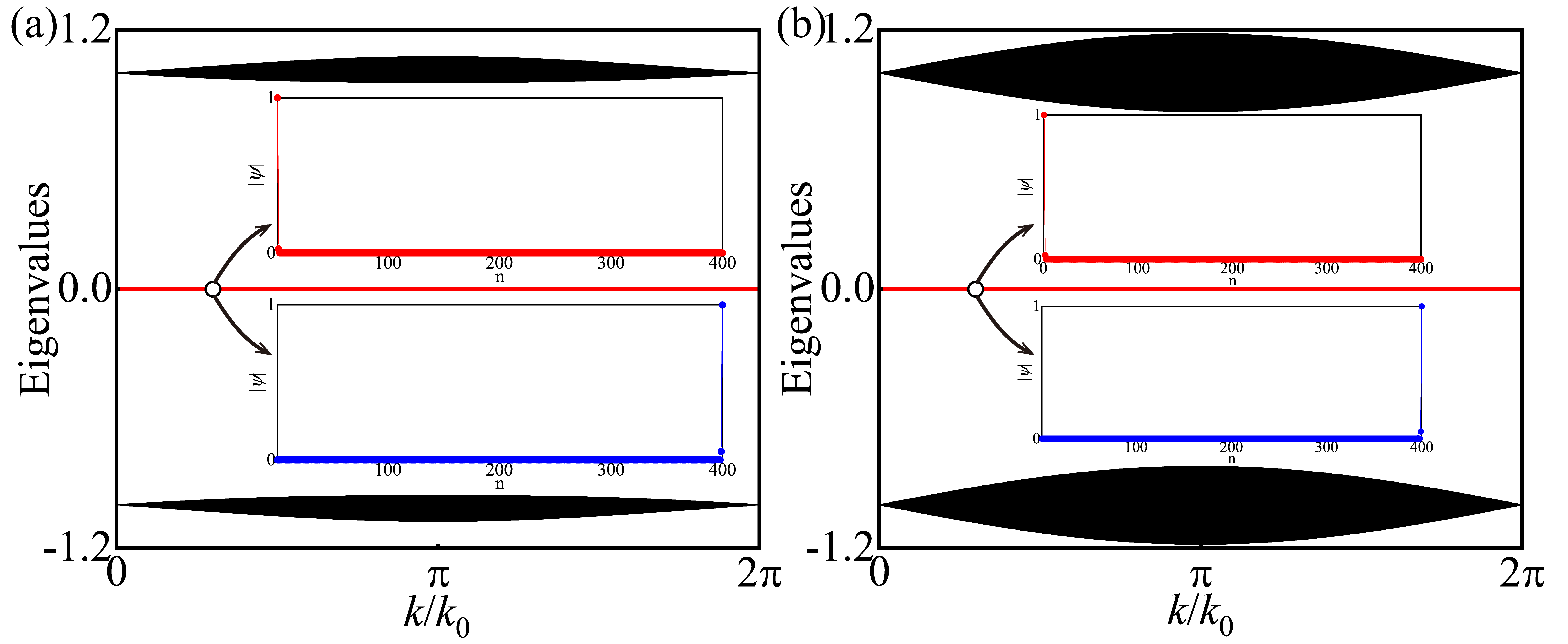}
    \caption{Eigenvalues of unperturbed Hamiltonian $H_{0}\left(k\right)$ for (a) QSHI phase with parameters $\left(\lambda_{\rm I}^{A},\lambda_{\rm I}^{B}\right)=\left(0.06t,0.03t\right)$ and (b) NLS phase with parameters $\left(\lambda_{\rm I}^{A},\lambda_{\rm I}^{B}\right)=\left(0.06t,-0.03t\right)$. The inserts show the absolute value of eigenvectors of the two-fold degenerate zero-energy flat band at $k/k_{0}=0.3\pi$ (empty circle).}
    \label{Fig8}
\end{figure}

In both the QSHI phase and NLS phase of the MKM model, the eigenvalue spectrum of the unperturbed term $H_{0}\left(k\right)$ includes a twofold-degenerate zero-energy flat band, as illustrated in Fig. \ref{Fig8}. Moreover, at fixed $k$, the eigenvectors of this flat band strongly localize at the endpoints of the two-leg ladder system, corresponding to the edge states at the lower boundary and upper boundary of the zigzag-terminated MKM model nanoribbon, as demonstrated by the inserts of Fig. \ref{Fig8}. For the edge states localized at the lower boundary, it is reasonable to propose a solution with zero amplitudes on sites ${\rm c}$ and ${\rm d}$ \cite{Rahmati2023}. By inserting this solution into the eigenfunction: $H_{0}\left(k\right)\ket{\psi_{\rm lower}\left(k\right)} = 0$, we derive:
\begin{equation}
    \left[
    \begin{array}{cc}
        -\lambda^{\prime} & t \\
        0 & \lambda
    \end{array}
    \right]
    \left[
    \begin{array}{c}
        \psi_{n}^{a} \\
        \psi_{n}^{b}
    \end{array}
    \right]
    +\left[
    \begin{array}{cc}
        -\lambda^{\prime} & 0 \\
        t & \lambda
    \end{array}
    \right]
    \left[
    \begin{array}{c}
        \psi_{n+1}^{a} \\
        \psi_{n+1}^{b}
    \end{array}
    \right]=0
    \label{EQA3}
\end{equation}
where $\psi_{n}^{a(b)}$ is the amplitude of the wavefunction on the ${\rm a}({\rm b})$ site. Here and in what follows, we drop the momentum dependence of the parameters $\lambda^{\prime}$ and $\lambda$ for notational brevity. Solving Eq. (\ref{EQA3}) yields
\begin{equation}
    \left[
    \begin{array}{c}
        \psi_{n+1}^{a}  \\
        \psi_{n+1}^{b}
    \end{array}
    \right]=\mathcal{M}^{n}\left[
    \begin{array}{c}
        \psi_{1}^{a}  \\
        \psi_{1}^{b}
    \end{array}
    \right],
    \label{EQA4}
\end{equation}
with
\begin{equation}
    \mathcal{M}=\left[
    \begin{array}{cc}
        -1 & \frac{t}{\lambda^{\prime}} \\
        \frac{t}{\lambda} & -\frac{t^{2}}{\lambda\lambda^{\prime}}-1
    \end{array}
    \right].  \notag
\end{equation}
The eigenvalues of $\mathcal{M}$ are
\begin{equation}
    \Lambda_{\pm}=-\left(1+\frac{1}{2}\frac{t^{2}}{\lambda\lambda^{\prime}}\right)\pm{\rm sgn}\left(\lambda\lambda^{\prime}\right)\frac{t^{2}}{\lambda\lambda^{\prime}}\sqrt{\frac{1}{4}+\frac{\lambda\lambda^{\prime}}{t^2}}.
    \label{EQA5}
\end{equation}

To achieve a convergent solution, it suffices to select the eigenvector for which $\left|\Lambda_{+(-)}\right|<1$, corresponding to the QSHI phase, we have $\left|\Lambda_{+}\right|<1$. Conversely, for ${\rm sgn}\left(\lambda\lambda^{\prime}\right)=-1$, indicating NLS phases, we have $\left|\Lambda_{-}\right|<1$. In both cases, the eigenvector components of $\mathcal{M}$ are:
\begin{equation}
    \left\{
    \begin{array}{c}
        \phi_{1}^{2}\left(k\right)=\frac{1}{\lambda^{\prime}}+\frac{2}{\lambda^{\prime}}\sqrt{\frac{1}{4}+\frac{\lambda\lambda^{\prime}}{t^{2}}}  \\
        \phi_{2}^{2}\left(k\right)=-\frac{1}{\lambda}+\frac{2}{\lambda}\sqrt{\frac{1}{4}+\frac{\lambda\lambda^{\prime}}{t^{2}}}
    \end{array}
    \right.
    \label{EQA6}
\end{equation}
and normalization factor
\begin{equation}
    C^{2}=\frac{1}{\lambda^{\prime}}+\frac{2}{\lambda^{\prime}}\sqrt{\frac{1}{4}+\frac{\lambda\lambda^{\prime}}{t^{2}}}-\frac{1}{\lambda}+\frac{2}{\lambda}\sqrt{\frac{1}{4}+\frac{\lambda\lambda^{\prime}}{t^{2}}}.
    \label{EQA7}
\end{equation}
Therefore, one can have $\psi_{1}^{a}\left(k\right)=\phi_{1}\left(k\right)/C$ and $\psi_{1}^{b}\left(k\right)=\phi_{2}\left(k\right)/C$.
Using first-order perturbation theory, the energy dispersion of edge states localized at lower boundary is
\begin{equation}
    E^{l}\left(k\right)=\bra{\psi^{l}\left(k\right)}H_{1}\left(k\right)\ket{\psi^{l}\left(k\right)}.
    \label{EQA8}
\end{equation}
Since the edge states are localized at outermost two lattice sites, only $\ket{\psi^{l}\left(k\right)}=\left[\psi_{1}^{a},\psi_{1}^{b}\right]^{T}$ contributes to the energy dispersion of the edge state. Therefore, the perturbed Hamiltonian is $H_{1}\left(k\right)=\left[\begin{array}{cc}
    \Delta^{\prime}\left(k\right) & t^{\prime}\left(k\right) \\
    t^{\prime}\left(k\right) & -\Delta\left(k\right)
\end{array}\right]$. After some algebra, the energy dispersion for lower boundary edge states is
\begin{equation}
    E^{l}\left(k\right)=\frac{12\cos{\frac{\sqrt{3}k a}{2}}}{C^{2}}.
    \label{EQA9}
\end{equation}
A similar process for the upper boundary gives rise to the energy dispersion
\begin{equation}
    E^{u}\left(k\right)=\frac{12\cos{\frac{\sqrt{3}k a}{2}}}{C^{\prime 2}},
    \label{EQA10}
\end{equation}
with $C^{\prime 2}=-\frac{1}{\lambda}-\frac{2}{\lambda}\sqrt{\frac{1}{4}+\frac{\lambda\lambda^{\prime}}{t^{2}}}-\frac{1}{\lambda^{\prime}}+\frac{2}{\lambda^{\prime}}\sqrt{\frac{1}{4}+\frac{\lambda\lambda^{\prime}}{t^{2}}}$. In the limit of $\lambda\lambda^{\prime}\ll 1$, the expansion $\sqrt{\frac{1}{4}+x}\sim\frac{1}{2}$ can be applied, causing $C^{2}=\frac{2}{\lambda^{\prime}}$ and $C^{\prime 2}=-\frac{2}{\lambda}$. Therefore, Eqs. (\ref{EQA9}) and (\ref{EQA10}) can be approximately expressed as
\begin{equation}
    E^{l}\left(k\right)=6\lambda^{\prime}\left(k\right)\cos{\frac{\sqrt{3}k a}{2}},
    \label{EQA11}
\end{equation}
and
\begin{equation}
    E^{u}\left(k\right)=-6\lambda\left(k\right)\cos{\frac{\sqrt{3}k a}{2}},
    \label{EQA12}
\end{equation}

\bibliography{reference}

\end{document}